# Breathing deformation model - application to multi-resolution abdominal MRI

Chompunuch Sarasaen, Soumick Chatterjee, Mario Breitkopf, Domenico Iuso, Georg Rose
and Oliver Speck *Otto von Guericke University (OVGU) Magdeburg, Germany*

*Abstract—* Dynamic MRI is a technique of acquiring a series of images continuously to follow the physiological changes over time. However, such fast imaging results in low resolution images. In this work, abdominal deformation model computed from dynamic low resolution images have been applied to high resolution image, acquired previously, to generate dynamic high resolution MRI. Dynamic low resolution images were simulated into different breathing phases (inhale and exhale). Then, the image registration between breathing time points was performed using the B-spline SyN deformable model and using cross-correlation as a similarity metric. The deformation model between different breathing phases were estimated from highly undersampled data. This deformation model was then applied to the high resolution images to obtain high resolution images of different breathing phases. The results indicated that the deformation model could be computed from relatively very low resolution images.

## I. INTRODUCTION

Magnetic Resonance Imaging (MRI) is a non-invasive medical imaging modality, applied in numerous diagnostic applications in radiology. MRI offers high spatial resolution for detecting disease and pathologic changes in tissue. Interventional MRI (iMRI) has advantages in term of non-ionizing radiation compared to Computed Tomography (CT) and high spatial resolution together with multi-contrast tissue properties compared to ultrasound. MRI-guided intervention differs from daily clinical MRI since a time series of images needs to be continuously acquired throughout the procedure, known as dynamic imaging. However, there has been a tradeoff between spatial and temporal resolution in such types of imaging. When the temporal resolution is critical, especially for dynamic liver imaging, spatial resolution is limited [1-2], the obtained time series images are considered as low resolution images compared to high resolution images in clinical routines. In practice, iMRI requires visualization of the lesion for percutaneous procedures. It is also challenged by breathing-related motion which can degrade image quality and lead to difficulties to delineate lesions [3-5].

In addition, there are several examinations reported the effect of motion in MRI as well as numerous studies aimed to make use of deformation motion model [6-8]. Nonetheless, there are several existing data available from previous scans, such as high resolution planning scans, but been neglected. Utilizing prior knowledge such as spatial resolution from high resolution images have not been explored.

The goal of this work is to overcome the limitation of spatial resolution in dynamic imaging by extracting deformation model related to breathing motion from fast low resolution time series data, with the assumption that breathing pattern should not change so much over time and then utilizing that deformation model by applying it on the available prior high resolution conventional scans to obtain high resolution, high quality dynamic images.

## II. METHODOLOGY

### A. Data acquisition and simulating dynamic imaging

Abdominal imaging data for this work was acquired on a SIEMENS MAGNETOM Skyra (3T) MR Scanner using T1 vibe sequence (TR: 3.64ms, TE: 1.45ms, Flip angle: 9.0 deg) with SPAIR fat suppression, distortion correction (2D) and pre-scan normalization. The field of view (FoV) of this acquisition was 384x216 mm, acquired 120 slices with a base resolution of 384x216 pixels (Voxel size: $1.0\times1.0\times1.6$ mm). All the slices were then cropped from the center to get only the abdomen to obtain the final resolution of 268x216x120 pixels, which has been considered as the high resolution image in this work. Dynamic imaging was simulated by acquiring the data in two different breathing positions (inhale and exhale), where the subject was asked to perform breath-hold in the particular breathing position. Image acquired in inhale has been considered as the time point 1 (TP1) and exhale as the time point 2 (TP2) of a dynamic imaging.

### B. Simulating low resolution data

Acquired data was artificially undersampled to simulate low resolution dataset, which is considered as the TP1 and TP2 of a low resolution dynamic imaging. Undersampling was performed by taking the center of the k-space (preserving the aspect ratio) of each slice and zero-filling the rest of the k-space. Undersampling was performed by taking 50% and 25% of the k-space, making the effective resolution as 190x153 pixels and 134x108 pixels respectively. To test the robustness and limitations of this approach, the data were highly undersampled by taking only 3%, 2%, 1% and 0.5% of the k-space (effective resolutions: 46x37, 38x31, 26x21, 18x15 respectively). Fig. 1 shows the comparison between 100% sampling mask (fully sampled) and examples of three undersampling masks, along with the resultant undersampled images for 50%, 3% and 0.5% undersampled.

### C. Image Registration

A 3D image registration toolbox ANTs (Advanced Normalization Tools) [9] was employed in order to build the deformation model. B-spline symmetric image normalization method (SyN) [10-11] was used for capturing the non-rigid deformation in abdominal organs, such as liver. In this work, cross-correlation was used as the similarity matric, spline

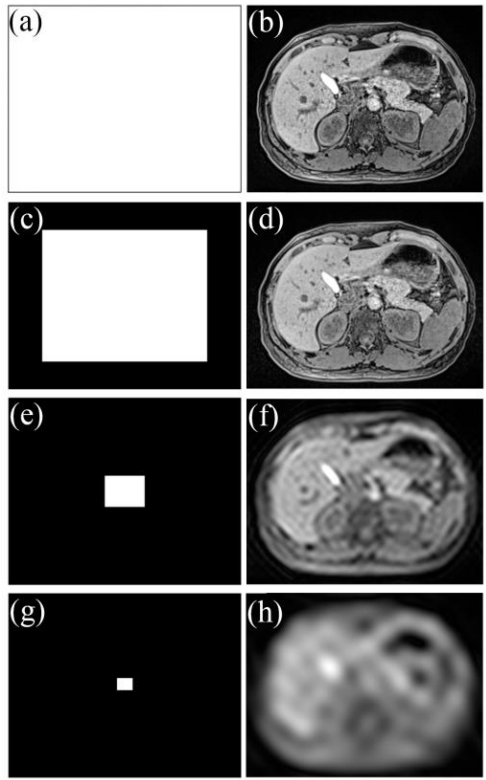

Figure 1. (a) and (b) 100% undersampling mask (i.e. fully sampled) and High resolution image, (c) and (d) 50% undersampling mask and undersampled image, (e) and (f) 3% undersampling mask and undersampled image , (g) and (h) 0.5% undersampling mask and undersampled image

distance was set to 40 mm, and the single cross-correlation was used as the similarity matric, spline distance was set to 40mm, and the single precision operation was performed to reduce computing overhead. During the registration, the undersampled inhale breathing phase (TP1) was selected as a moving image, while undersampled exhale phase (TP2) was selected as the fixed image.

### D. Application of deformation in multi-resolution

Deformation field extracted by performing registration on undersampled images, was merged with the high resolution TP1 using the B-Spline interpolation, to obtain high resolution TP2, so-called the deformed image. The resultant deformed image has then been compared with the real high resolution TP2 to measure the performance of this approach.

### III. RESULTS AND DISCUSSIONS

For performance comparison of the undersampled registrations, an additional deformation model was extracted by registering high resolution TP1 and TP2 (also referred as 100% of k-Space), and then that deformation model was applied on the high resolution TP1 to obtain the deformed image for 100% of k-Space. As no undersampling was involved, this results in the best possible output from this proposed approach. The deformation model obtained from this were then been treated as the ground-truth model.

### A. Deformation field and comparisons

The deformation field shows the movement of voxels from TP1 to TP2 in different planes (i.e. x,y,z). The deformation field obtained using ANTs, as shown in Fig. 2, shows how much each voxel has moved in each plane.

The deformation models obtained from registering various levels of undersampling were then compared against this ground-truth model based on the Hausdorff distance between them and in terms of percentage of errors, as shown in Fig. 3. It was observed that the more the level of undersampling were used for image registration, more error it contains. But the resultant deformed images are still very close to the actual high resolution TP2.

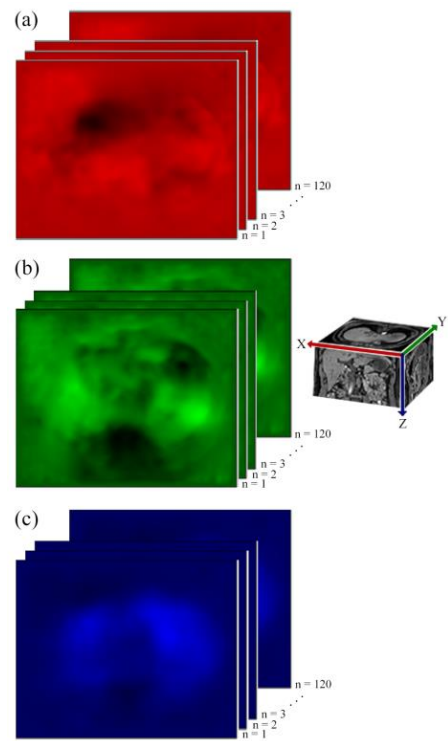

Figure 2. An example of deformation field obtained by registering undersampled image (0.5% of the k-Space) in (a) x-plane, (b) y-plane and (c) z-plane

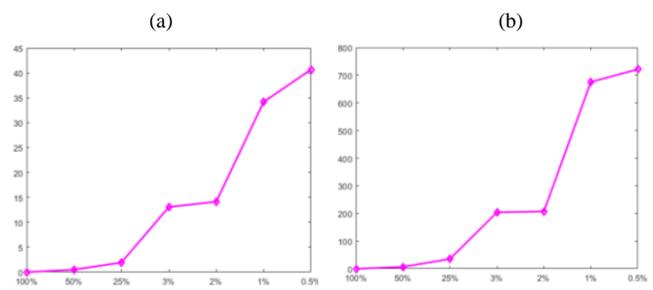

Figure 3. Comparison of the deformation fields obtained with various resolutions, against the deformation field obtained using high-resolution image, using (a) Error Percentage (b) Hausdorff Distance

## B. Comparison of the deformed images

The deformed images from various levels of undersampling were compared with the actual high resolution TP2 image using various similarity measures, such as - Pearson correlation coefficient, percentage of error, mean-squared error and Structural Similarity Index (SSIM), as shown in Fig. 4. It was observed for all of the mentioned similarity matrices, that the resultant deformed images obtained by registering 50% or 25% undersampled data, are very similar to the one obtained by registering full-sampled data. When more highly undersampled data (3%, 2% and so on) were used for registration, the quality of the resultant deformed image decreases, but still quite similar to the actual high resolution TP2, as shown in Fig. 5. The highest undersampled data that was used for registration was 0.5% of the original k-Space, and as observed, this approach still works for such a highly undersampled data.

## C. Discussions

Image registration is a time-consuming process. For undersampled data, the image registration should take less time as it contains less number of pixels and also should decrease the computation overhead. However, currently in this research work the k-Space were zero-filled during undersampling, in order to preserve the pixel resolution of the images. As the undersampled data used in this work had the same pixel resolution as the fully-sampled data, the image registration process took almost the same time as registering fully-sampled data. In future, this approach will be tested without zero-filling the k-space while undersampling, reducing the actual pixel size.

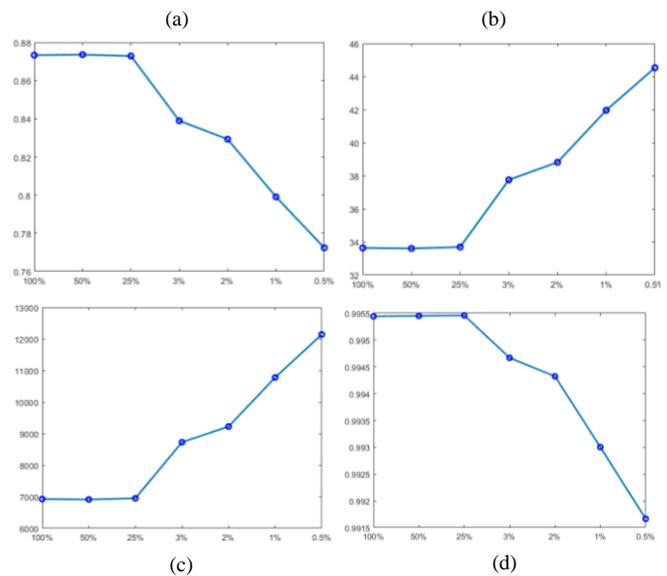

Figure 4. Comparison of the Deformed Image with the actual high resolution Time Point 2 using various similarity measures
(a) Pearson correlation coefficient (b) Error Percentage
(c) Mean-squared Error (d) SSIM

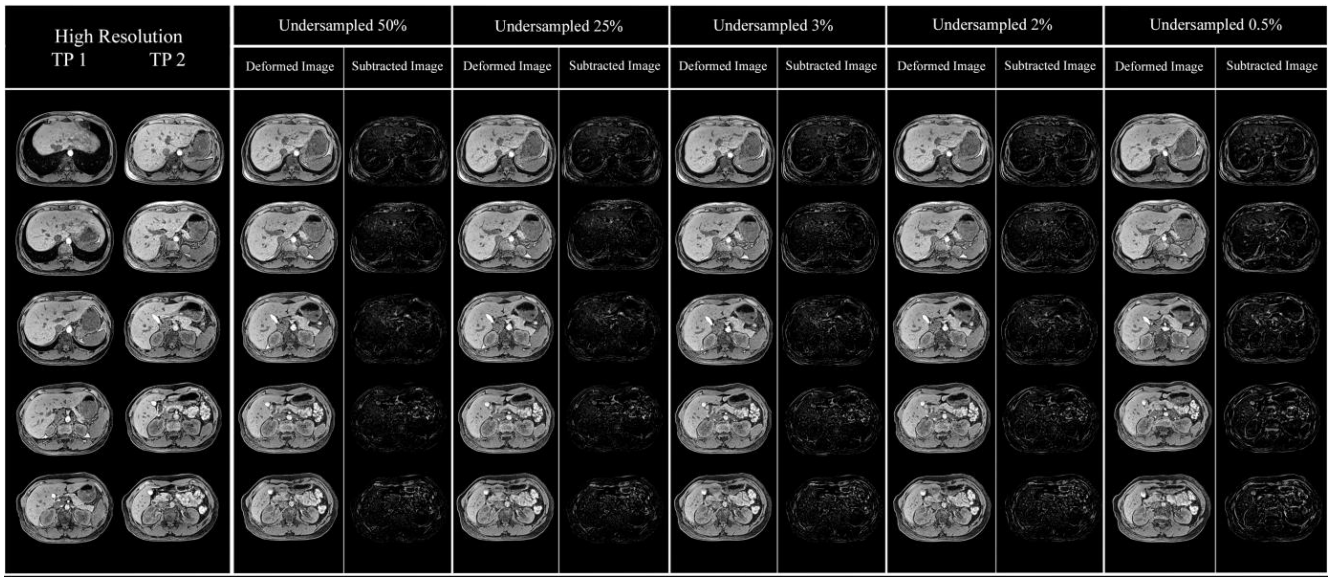

Figure 5. Sample output (Deformed Image) of the proposed approach for different levels of undersampling (i.e. 50%, 25%, 3%, 2%, 0.5% of the fully sampled k-space), along with the substraction (Substracted Image) of the output from the High Resolution time point 2 (TP2)

## IV. CONCLUSION

This proposed approach can make use of deformation models extracted from highly undersampled data. This could be helpful for acquiring dynamic images which can offer both high spatial and temporal resolution. The lowest effective resolution that this approach has been tested to be working with is 18x15 pixel (0.5% of the k-space of the high resolution image used for testing). The approach is yet to be tested for undersampling without zero-filling the k-Space, to speed up the registration process. The clinical feasibility of this approach have also to be investigated.

## ACKNOWLEDGMENT

This work was conducted within the context of the International Graduate School MEMoRIAL at Otto von Guericke University (OVGU) Magdeburg, Germany, kindly supported by the European Structural and Investment Funds (ESF) under the programme "Sachsen-Anhalt WISSENSCHAFT Internationalisierung" (project no. ZS/2016/08/80646). The experimental procedures involving human subjects described in this paper were approved by the Institutional Review Board.